\documentclass[useAMS]{mn2e}
\usepackage{epsfig}
\newcounter{newone}
\newcommand{\pap}{Paper \Roman{newone}} 
\def\mum{$\mu$m }

\begin{document}

\title[Confusion and Gravitational Lensing Limits for the SZ Increment]
{Estimates of Confusion and Gravitational Lensing Limits in
Sunyaev--Zel'dovich Increment Measurements}

\author[Zemcov, Newbury \& Halpern]{
\parbox[t]{\textwidth}{
\vspace{-1.0cm}
Michael Zemcov$^{1}$, 
Peter Newbury$^{1}$, 
Mark Halpern$^{1}$
}
\vspace*{6pt}\\
$^{1}$ Department of Physics \& Astronomy, University of British Columbia,
       Vancouver, BC, Canada \\
\vspace*{-0.5cm}}

\date{Draft 2003 February 22}

\maketitle

\begin{abstract}
We present estimates of the confusion limit of measurements of the
Sunyaev--Zel'dovich effect (SZ) increment near its peak intensity
based on Monte Carlo simulations of sub-millimetre observations of
galaxy clusters. Specifically, we evaluate the contribution of
gravitational lensing of high redshift background sources by a target
cluster to its inferred SZ signal.  We find that the background
confusion limit is $0.6$ mJy per beam without any lensing.  With
gravitational lensing, the confusion limit increases to $0.9$ mJy per
beam.  Removing bright sources in the gravitationally lensed fields
decreases the confusion limit to $0.7$ mJy per beam, essentially
background levels.  These limits are found by removing exceptionally
poor fits to the expected SZ profile, which are due to the different
structure of the SZ and lensing angular profiles.

We conclude that experiments designed to measure the SZ increment must
have high enough angular resolution to identify background point
sources while surveying enough of the sky to determine the SZ profile.
Observations of the SZ increment with SCUBA, which has sufficient
sensitivity and resolution for the task, should be straight forward.
\end{abstract}

\begin{keywords}
galaxies: clusters: general -- cosmic microwave background --
cosmology: gravitational lensing -- methods: numerical -- methods:
statistical 
\vspace*{-0.5cm}
\end{keywords}

\section{Introduction}

The Sunyaev--Zel'dovich (SZ) effect is a spectral distortion of the
cosmic microwave background (CMB) radiation.  It is caused by the
interaction of CMB photons and highly energetic electrons in a plasma.
Usually, this plasma is the diffuse $T_{\mathrm{e}} = 10^{7}$ K gas
present near the center of a galaxy cluster.  On average, CMB photons
are scattered to shorter wavelengths in collisions with relativistic
electrons, producing a decrement in apparent CMB temperature at long
wavelengths and an increment at short wavelengths.  Measurements of
the SZ increment, particularly at its $850$ \mum peak, will be a very
important probe of the Universe's large scale structure in the future
(e.g. Birkinshaw (1999)).  However, many diverse effects conspire to
make measurement of the sub-millimetre (sub-mm) SZ increment
difficult, including the brightness of the sub-mm sky, the need for
stable, accurate instrumentation, and the small amplitude of the SZ
increment signal.

SZ increment measurements are made more challenging still because of
other sources of flux in a cluster field.  These miscellaneous flux
sources include the Butcher--Oemler effect, the Rees--Sciama effect,
and gravitational lensing of primordial CMB anisotropies. Blain (1998)
gives a detailed discussion of these sources of flux in galaxy
clusters, and comes to the conclusion that none of these effects
produces a signal strong enough to compete with the SZ increment in
galaxy clusters.

However, source confusion can play a large part in increasing the
danger of erroneous measurements.  The term source confusion is used
to describe the variance in flux observed in a map due to unresolved,
faint sources in the field, the level of which depends on the
resolution of the telescope and the background source number count in
the observational passband.  While it is true that sub-mm emission
from within an X-ray cluster is rare, background sources at higher
redshift, $z$, are common and relatively bright. This means that
source confusion is inevitable in a cluster field, and may mimic the
SZ effect.  The effect of source confusion may be compounded many
times by gravitational lensing, which brightens high $z$ background
sources behind a cluster.  This paper evaluates the effect of
gravitational lensing and its contribution to source confusion in
sub-mm measurements of the SZ effect.

We are engaged in a program to measure the SZ increment in X-ray
bright clusters with the Sub-millimetre Common User Bolometer Array
(SCUBA) on the James Clerk Maxwell Telescope (JCMT).  We wish to
determine the confusion limit of these measurements so that we can
design a statistically valid observing program, and additionally to
determine if gravitational lensing of background sources by the target
cluster will worsen the confusion limit or bias our estimate of the
size of the effect.  A preliminary description of the measurements
themselves is found in Zemcov et al. (2002), hereafter \pap.  We have
performed Monte Carlo simulations of SCUBA observations of galaxy
clusters including lensing and a realistic source model.  Blain (1998)
has studied this problem previously, but was forced to model galaxy
formation and evolution using optical and infra-red source counts
since observationally determined sub-mm background source counts were
not available.  Further, Blain (1998) considers only single pixel
measurements of the SZ effect while we use the high resolution
available with SCUBA to identify point sources and remove them from
analysis.  A standard $(\Omega_{M} = 0.3, \Omega_{\Lambda} = 0.7)$
cosmology is assumed (Bennett et al. (2003)), and all errors are
1-$\sigma$ unless otherwise stated.

\section{Experiment}

\subsection{Lensed Sub-mm Map Simulations}

In order to realistically simulate the effect of gravitational lensing
on source confusion, accurate source counts are required. Borys et
al. (2002) have analyzed sub-mm HDF data and conclude that the number
of sources with amplitudes greater than flux $S$, $N(>S)$, is

\begin{equation}
N(>S) = N_{0} \ \left(\frac{S}{S_{0}} \right)^{-\alpha} \left(1 +
\frac{S}{S_{0}} \right)^{-\beta}
\end{equation}   

\noindent with $S_{0} = 10$ mJy, $\alpha = 0.8$, $\beta = 2.5$, and
$N_{0} = 1.55 \times 10^{3}$ deg$^{-2}$.  We assume that the SCUBA
background sources are at most weakly clustered on scales relevant to
this experiment.  This means that a Poisson distribution of sources
uniformly and randomly distributed in a field is fully consistent with
the data.

Because equation (1) yields cumulative counts, we take its numerical
derivative.  For each amplitude bin, a Poisson distribution with mean
equal to the number of sources in that bin is used to give a random
number of sources for each amplitude bin consistient with the source
counts. A random number from a uniform distribution ranging between
the limits of the relevant amplitude bin is added to each of the
source amplitudes to randomize the source strengths. This procedure
produces an ensemble of random source strengths in the range $0.01$
mJy to $25$ mJy. Each point source amplitude is then associated with a
uniformly distributed random position in the field. The simulated map
created in this way is referred to as a `background' field and is
statistically the same as a sub-mm field observed at random in the
sky.  Because SCUBA sources are all likely to be background objects,
they will all be lensed to some extent when observed in the region of
a foreground cluster.

Our gravitational lensing algorithm is based on that used in Newbury
\& Spiteri (2002)\footnote{A general realization of this algorithm can
be found at {\tt
http://www.astro.ubc.ca/people/newbury/siam/lens.html}.  Here we have
used a version restricted to circular lenses.}, and the parameteric
lensing model utilized in this experiment is the singular isothermal
sphere (SIS).  If $\mathbfit{y}$ is the position of the point source
in the source plane in arcsec, $\mathbfit{x}$ is the position of the
lensed image in the deflector plane in arcsec, $D_{\mathrm{ds}}$ is
the distance between the source and the deflector, $D_{\mathrm{s}}$ is
the distance to the source, and $\sigma$ is the velocity dispersion of
the deflector in km s$^{-1}$, then the lensing model is

\begin{equation}
\mathbfit{y} = \mathbfit{x} - 28.8 \; \frac{D_{\mathrm{ds}}}{D_{\mathrm{s}}}
\left( \frac{\sigma}{1000} \right)^{2} \frac{\mathbfit{x}}{|\mathbfit{x}|}.
\end{equation}

\noindent The numerical values $28.8$ and $1000$ have the units arcsec
and km s$^{-1}$ respectively.  Note that because the SIS model is
spherically symmetric, changes in the apparent position of the sources
must be radial.  The distances are comoving distances determined via
integration over redshift using the standard cosmology mentioned in
the Introduction.  The values of the parameters $z_{\mathrm{d}}$ and
$\sigma$ are chosen to be the experimentally determined values for Cl
$0016+16$ (see Table $1$), as it is one of the clusters studied in
\pap. Although a relatively simple model, the SIS quite accurately
parametrizes gravitational lensing by spherical mass distributions.

The distance parameters in equation ($2$) are computed using
$z_{\mathrm{s}} = 2.5$ in our simulations.  We can assume all sources
are in a plane at one redshift because the $z$ dependence of the lens
equation and the effect of cosmological dimming approximately cancel
above $z \simeq 1$, effectively producing a $2$ dimensional lensing
background.  The redshifts of the background SCUBA sources are not
known, and so an estimate for $\bar{z}$ must be made. It is thought
that they reside in the range $1 < z < 4$, with a peak number density
between $z = 2$ and $z = 3$. Because the increase due to gravitational
lensing is approximately constant above $z \simeq 1$ and the sources'
redshifts are not known, simulations using a constant source redshift
of $2.5$ are the same as those using a redshift of $3.5$ or $4.5$.

\begin{center}
\begin{tabular}{lll}
\multicolumn{3}{c}{Table 1: Cl $0016+16$ Parameters} \\
\hline 
 Parameter          & Value              & Reference \\
 $z_{\mathrm{d}}$   & $0.55$             & Reese et al. (2000) \\
 $\sigma$           & $1324$ km s$^{-1}$ & Smail et al. (1995) \\ 
\hline
\end{tabular}
\end{center}

As well as shifting an image's location, lensing amplifies the flux of
the background sources.  The SIS is used to analytically determine
radial change and amplification as a function of initial position for
each source in the background field; the map produced in this way is
referred to as a `lensed' map.

To simulate the telescope's response to a group of point sources
properly, the lensed map is convolved with a Gaussian beam with the
same FWHM as the JCMT. If desired, an SZ effect signal can be inserted
into the map at this point. Additionally, point sources greater than a
given flux limit may be removed from the field at this stage. This map
is referred to as a `convolved' field, or as a `convolved and cleaned'
field.

\subsection{SZ Effect Amplitude Determination}

SCUBA measurements are differenced in order to minimize the effects of
emission from the atmosphere and the telescope itself.  In order to
properly simulate a differential instrument's response to the sky, it
is necessary for simulated maps to be sampled in the same way as the
sky is in an actual measurement.  Real SCUBA pointing data (taken from
a $3$ hour integration in \pap) is played across the convolved
field.  Random noise equal to that of a real observation is added to
the time stream; data from bolometers made noisy by this process are
removed from subsequent analysis.

A matrix inversion method is used to produce a seven pixel `map' of
the data (this method is discussed in Wright, Hinshaw \& Bennett
(1996)). In this map, the $i$th pixel gives the average value of the
map in an annulus with outer radius $40 i$ arcsec and inner radius $40
(i - 1)$ arcsec.  This binning can be applied to SZ data as the SZ
effect intensity distribution is spherically symmetric to first order.
A singuar value decomposition (SVD) method is used to invert the
matrix (Press et al. (1992)).  An example of data points determined
via this method is shown in Fig.~$1$.

Commonly, an isothermal $\beta$ model, given by equation ($3$), is
used to describe SZ effect intensity distributions.  If $I$ is the SZ
increment flux at angular distance $\theta$ from the centre of the
cluster, $I_{0}$ is the central increment value, and $\theta_{c}$ and
$\beta$ are parameters characterizing the cluster, then

\begin{equation}
I = I_{0} \; \left(1 + \frac{\theta^{2}}{\theta_{c}^{2}} \right)^{(1 -
3 \beta)/2}.
\end{equation}

\noindent In our simulations, the characteristic values are taken to be
$\theta_{c} = 50$ arcsec and $\beta = 0.75$. These are representative
of values one finds in observations of clusters (e.g.~Reese et
al. (2000)).  A flux density, $I_{0}$, which corresponds to the flux
of a point source in the central bolometer is recorded.  A generalized
$\chi^{2}$ must be used to characterize the goodness of fit, as the
data points determined via SVD are heavily correlated. If
$x_{\mathrm{m}}$ are the model points, $x_{\mathrm{d}}$ are the data
points, and $C_{\mathrm{ij}}$ is the correlation matrix found in the
SVD, then

\begin{equation}
\chi^{2} = (x_{\mathrm{m}} - x_{\mathrm{d}})_{\mathrm{i}} \
C_{\mathrm{ij}}^{-1} \ (x_{\mathrm{m}} -
x_{\mathrm{d}})_{\mathrm{j}}.
\end{equation} 

The isothermal $\beta$ profile is fit to the data points and the
amplitude $I_{0}$ and $\chi^{2}$ of the fit is recorded.  The
amplitude information allows us to understand the source confusion we
expect, while the $\chi^{2}$ statistic is a gauge of how well the
isothermal $\beta$ profile describes our data. A good fit is
characterized by a $\chi^{2}$ approximately equal to the number of
degrees of freedom in our fit, namely $5$.  An example of a isothermal
$\beta$ profile fit to simulated data is shown in Fig.~$1$.  Because
the data set consists of differences, it is insensitive to the mean
brightness. Fitting an isothermal $\beta$ profile is therefore a two
parameter fit for the baseline and amplitude.

\begin{center}
\begin{figure}
\centering
\epsfig{file=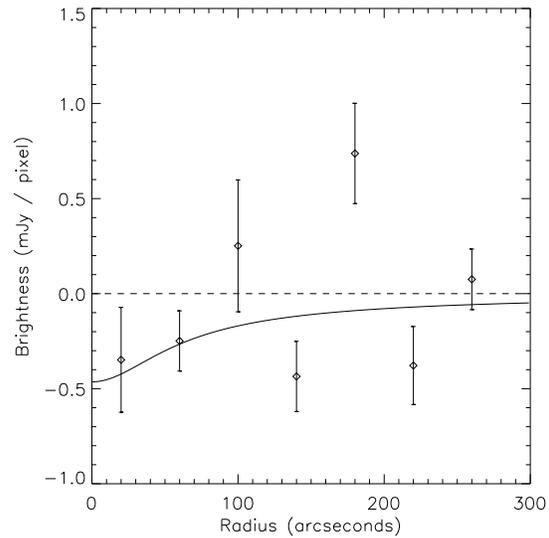,width=0.4\textwidth}
\caption{The results of a typical simulation described in Section
$2.2$. The points are averages over annular rings with $40$ arcsec
widths found via a matrix inversion method. The solid line below the
dashed zero level is the best fit isothermal $\beta$ profile given in
equation (3). The fit happens to be negative in this simulation, but
in general can have either sign.}
\label{fig:1}
\end{figure}
\end{center}

\subsection{Confusion Limit Determination}

This simulation process is repeated $N$ times to build a statistical
understanding of the source confusion we expect in a typical
observation of a cluster.  Fig.~$2$A shows the best fit SZ amplitude
and $\chi^2$ for $N = 100$ simulations of observations of unlensed
background fields.

Fig.~$2$B shows the inferred SZ amplitudes obtained in $N = 400$
simulations of observations of gravitationally lensed fields produced
as described above.  From these figures, it can be observed that
gravitationally lensed cluster fields often yield larger SZ fits than
unlensed fields do.  However, a very large $\chi^{2}$ accompanies a
gravitationally lensed field's larger amplitude, vastly increasing the
likelihood that the fit would be rejected. This is because the lensed
source appears at the Einstein ring, not at the cluster centre.

Fig.~$2$C shows $N = 100$ simulations of observations of
gravitationally lensed fields with sources greater than $6$ mJy
removed.  This is approximately the 3-$\sigma$ level for one night's
observing.  In order to avoid biasing our analysis scheme, we remove
sources greater than $6$ mJy in the target beam, and greater than $12$
mJy in the off-source beam.  In the SCUBA double difference there is
one target position and two off-source positions in any given
observation, with the array integrating on each off-source position
for half as long as it does on the target position.  Thus, a point
source in either of the off-source positions will be reported as
having negative one half its amplitude in the reconstructed image.
This means that a negative $6$ mJy point source in a map actually has
a positive $12$ mJy amplitude on the sky, and could have originated in
\textit{either} off-source field of view.  Notice that removing the
3-$\sigma$ detections essentially removes all the high $\chi^2$ fits
from the population.

\begin{figure}
\centering
\epsfig{file=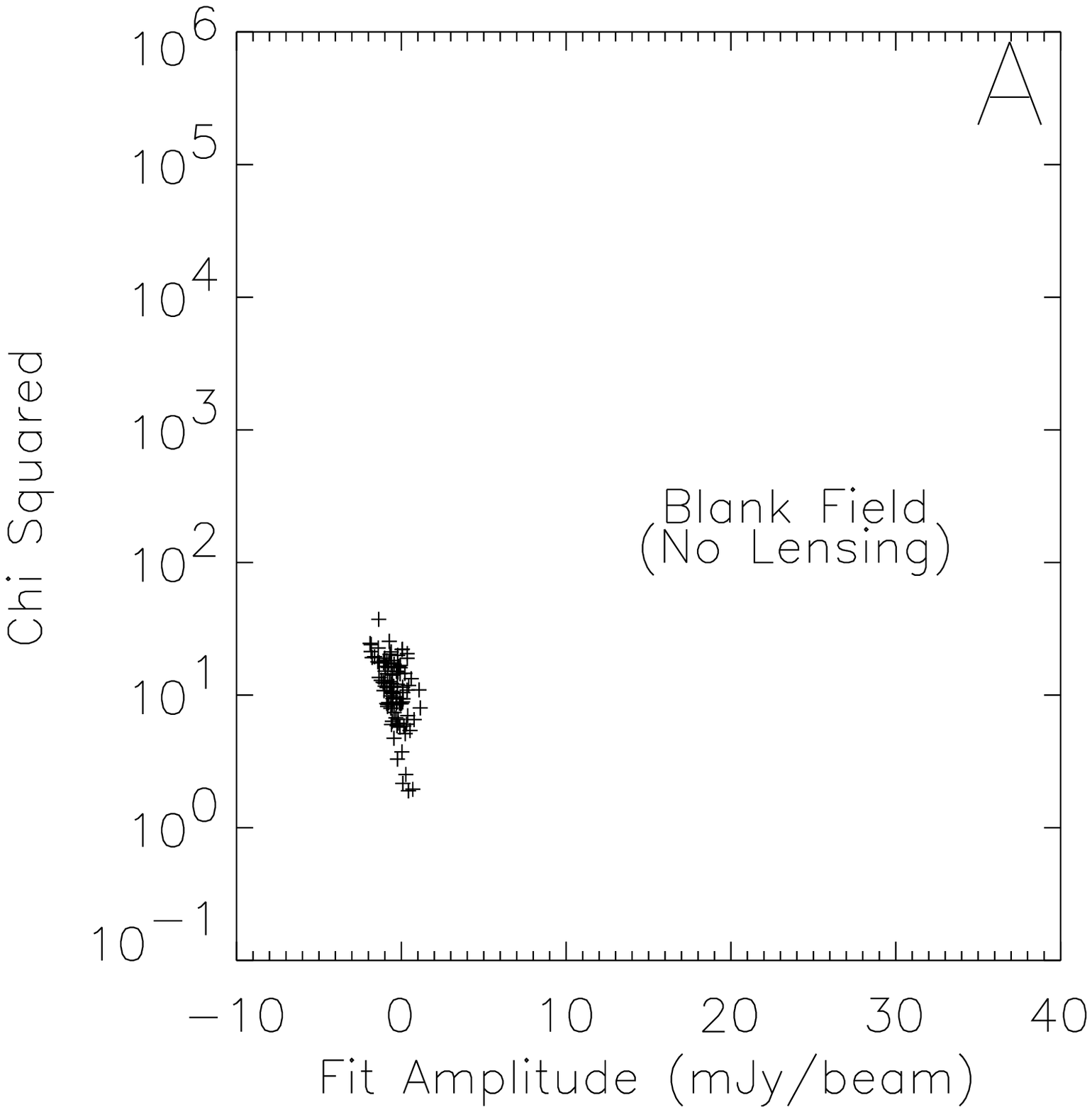,width=0.35\textwidth} 
\newline
\epsfig{file=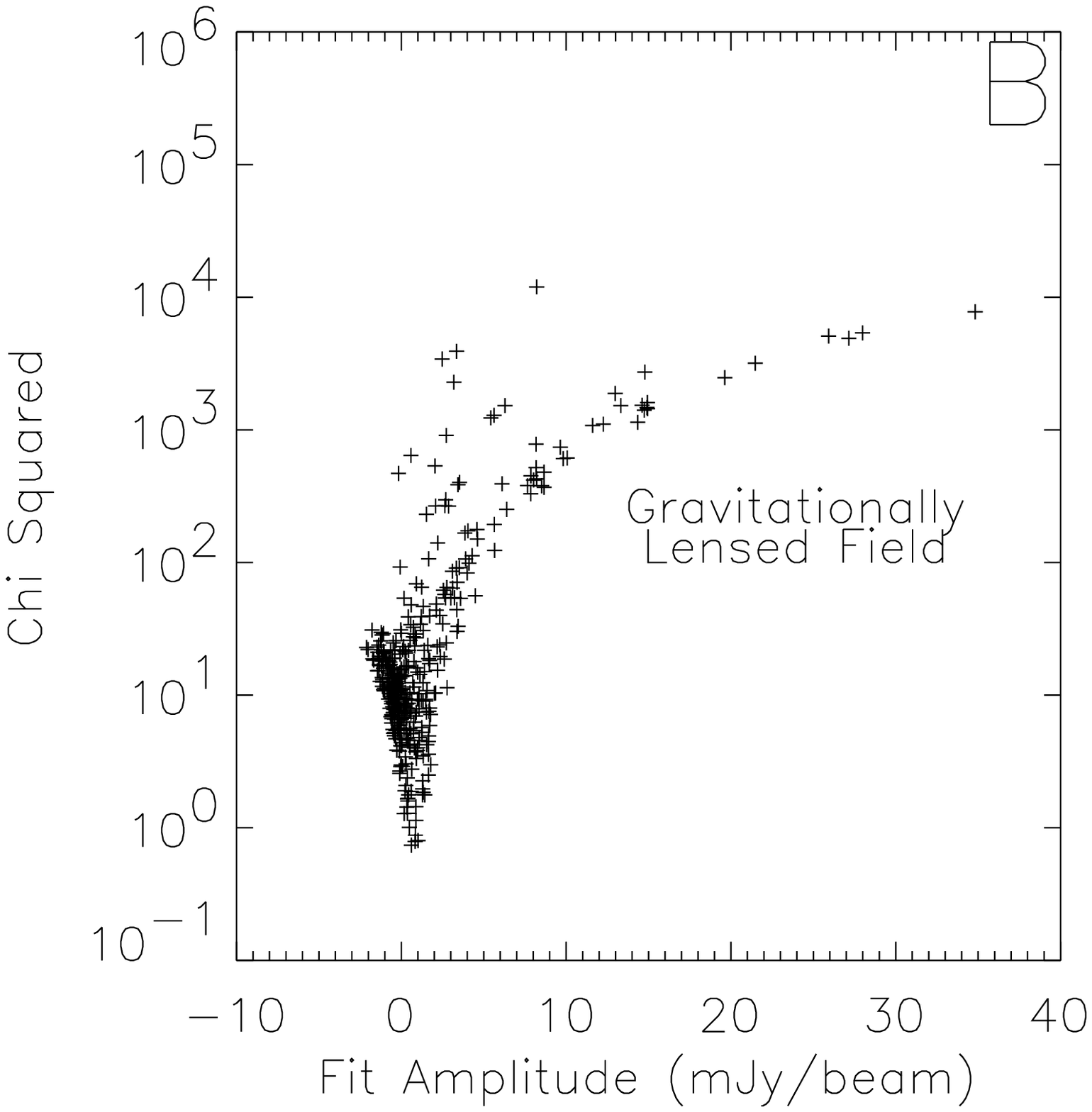,width=0.35\textwidth} 
\newline
\epsfig{file=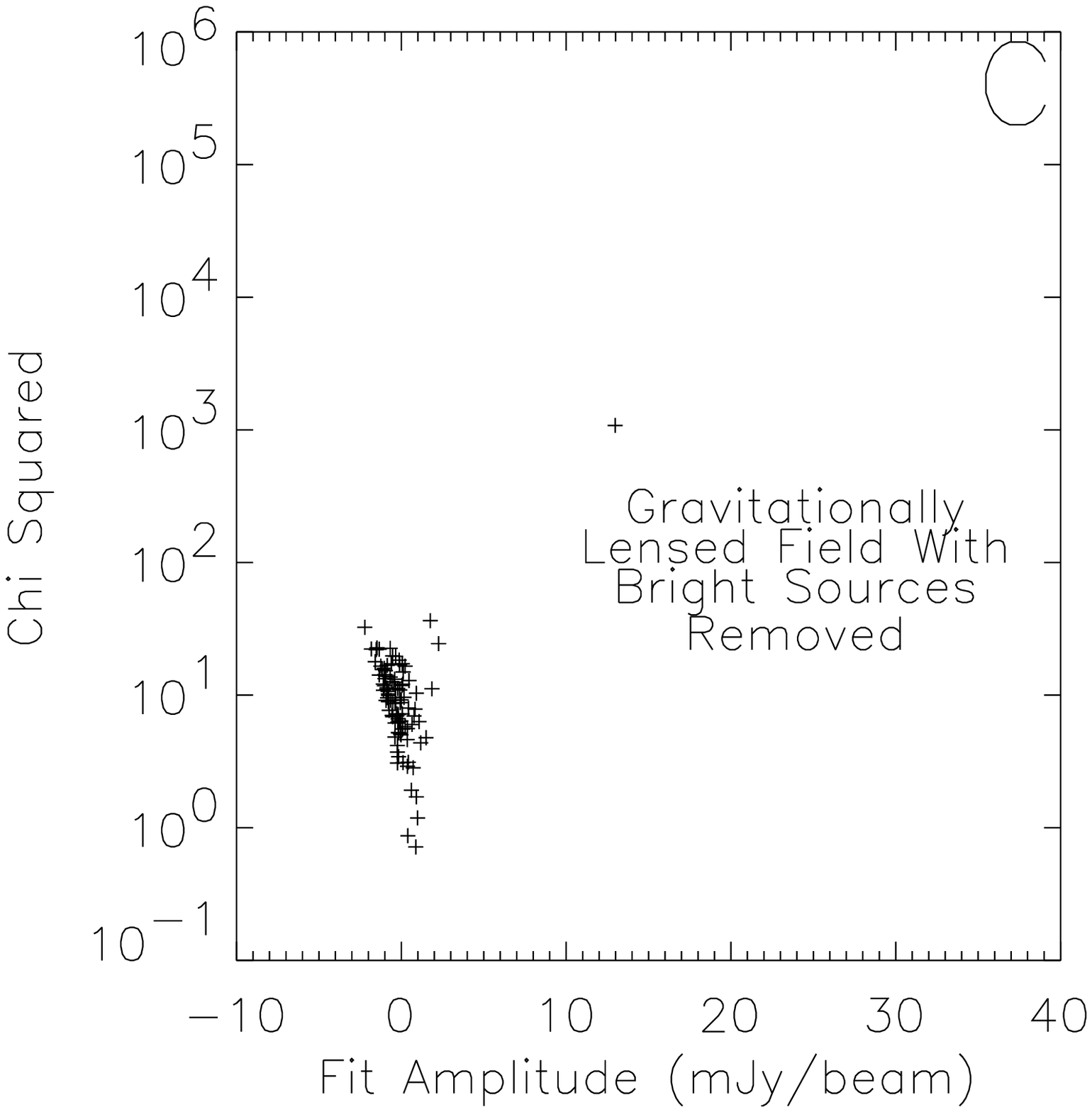,width=0.35\textwidth}
\newline
\caption{The results of our confusion limit simulations. Each of these
plots shows $\chi^{2}$ as a function of isothermal $\beta$ profile
amplitude for $N$ simulations. Panel A is the data from $N = 100$
simulations of background fields. Panel B shows the data for $N = 400$
simulations of gravitationally lensed fields with no bright sources
removed. The high value of $N$ is necessary to give a good sampling as
the distribution is sparse in the gravitationally lensed case. Panel C
shows the data from $N = 100$ simulations of gravitationally lensed
fields with point sources greater than $6$ mJy in the `on source' beam
and greater than $12$ mJy in the `chopped' beam removed.  These plots
show that gravitational lensing and the SZ effect have different
radial profiles, and fits where large lensed objects are present in
the field are poorly described by the data.}
\label{fig:2}
\end{figure}

Fig.~$3$ shows histograms of the $I_{0}$ from Figure $2$ in the
three cases.  The standard deviations of the distributions shown in
the three panels of Fig.~$3$ are $\sigma_{\mathrm{A}} = 0.63$ mJy
per beam, $\sigma_{\mathrm{B}} = 36.9$ mJy per beam and
$\sigma_{\mathrm{C}} = 1.54$ mJy per beam. These do not characterize
the widths of these non-Gaussian distributions well.  We choose to
define the confusion limit to be the standard deviation of these
distributions after excluding those fits for which $\chi^2 \geq 20$.
This $\chi^2$ cut removes 9\% and 11\% of distributions A and C.  We
calculate $\chi^2$ based on the detector noise, whereas the variance
due to confusion is $2.3$ times larger than detector noise.
Therefore, we expect this cut to remove 10\% of the sources in these
two nearly Gaussian distributions.  However, 140 of the 400
simulations in panel B are removed by the $\chi^2$ cut since that
distribution has a large tail of positive detections.  The resulting
confusion limits are $0.58$ mJy per beam for unlensed fields, $0.85$
mJy per beam for lensed clusters with poor $\chi^2$ samples removed,
and $0.66$ mJy per beam for lensed clusters with 3-$\sigma$ sources
removed.  The means of the three distributions in Fig.~$3$, after
the $\chi^2$ cut, are all consistent with zero.

\begin{figure}
\centering
\epsfig{file=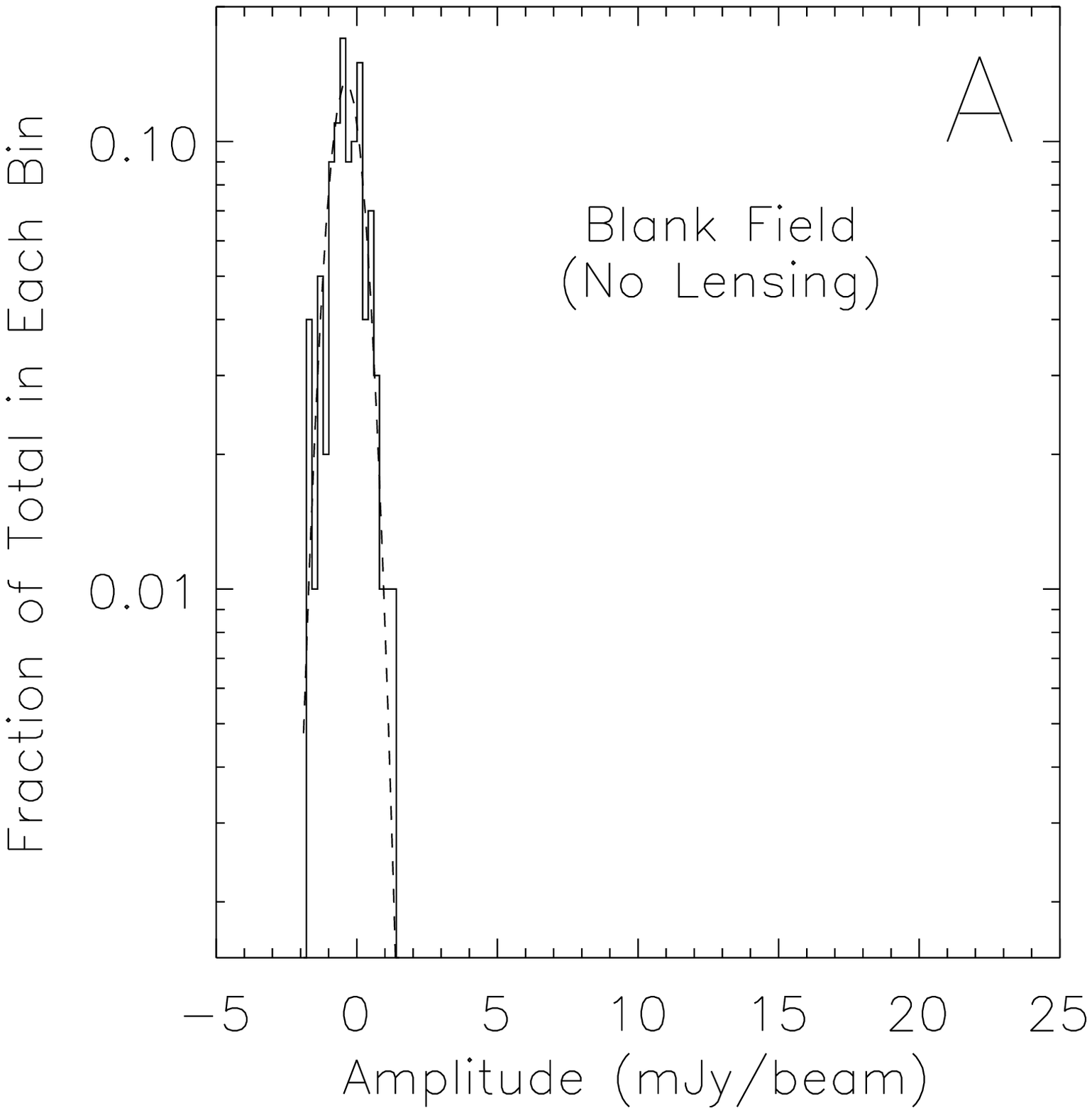,width=0.35\textwidth}
\newline
\epsfig{file=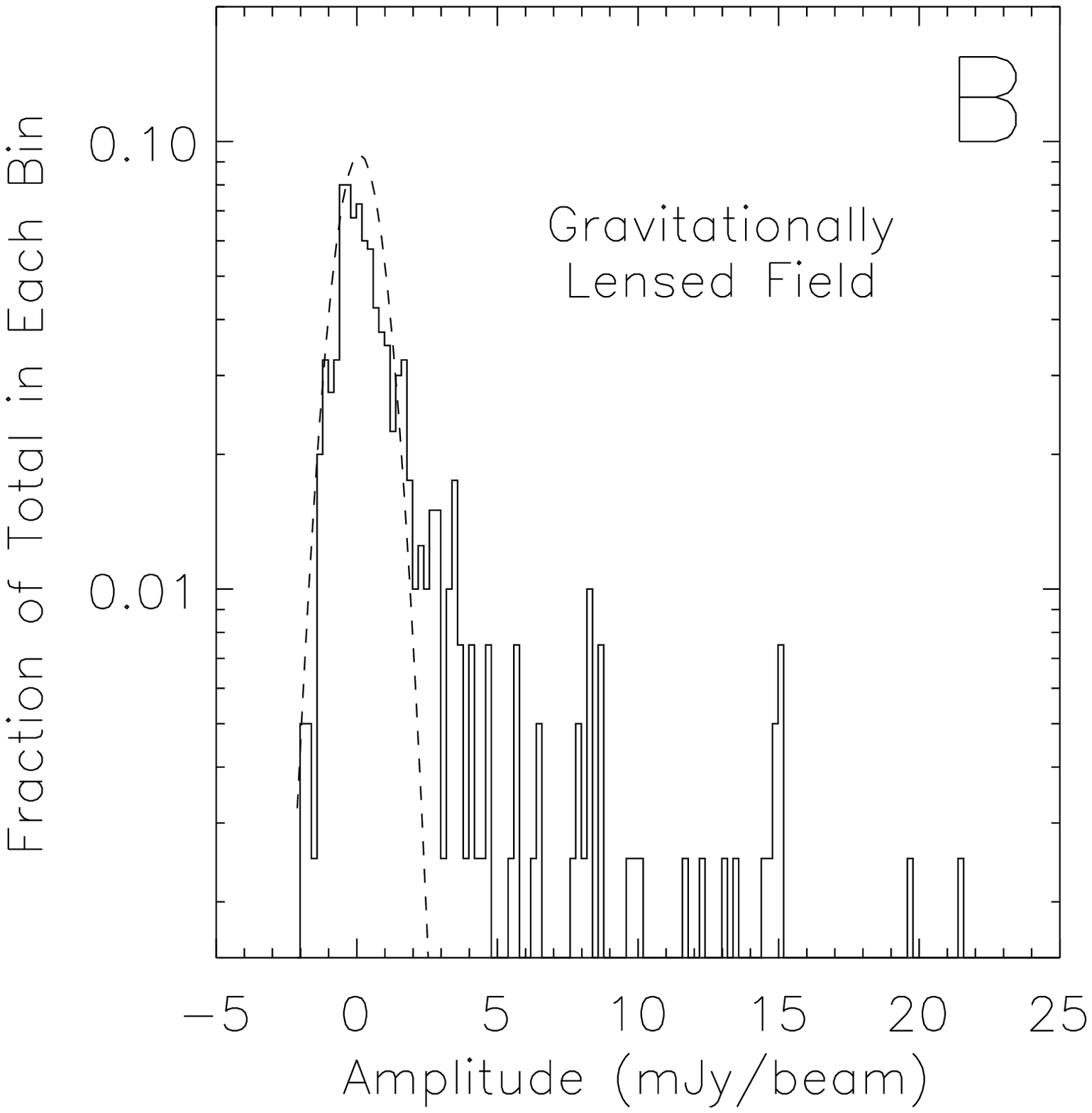,width=0.35\textwidth}
\newline
\epsfig{file=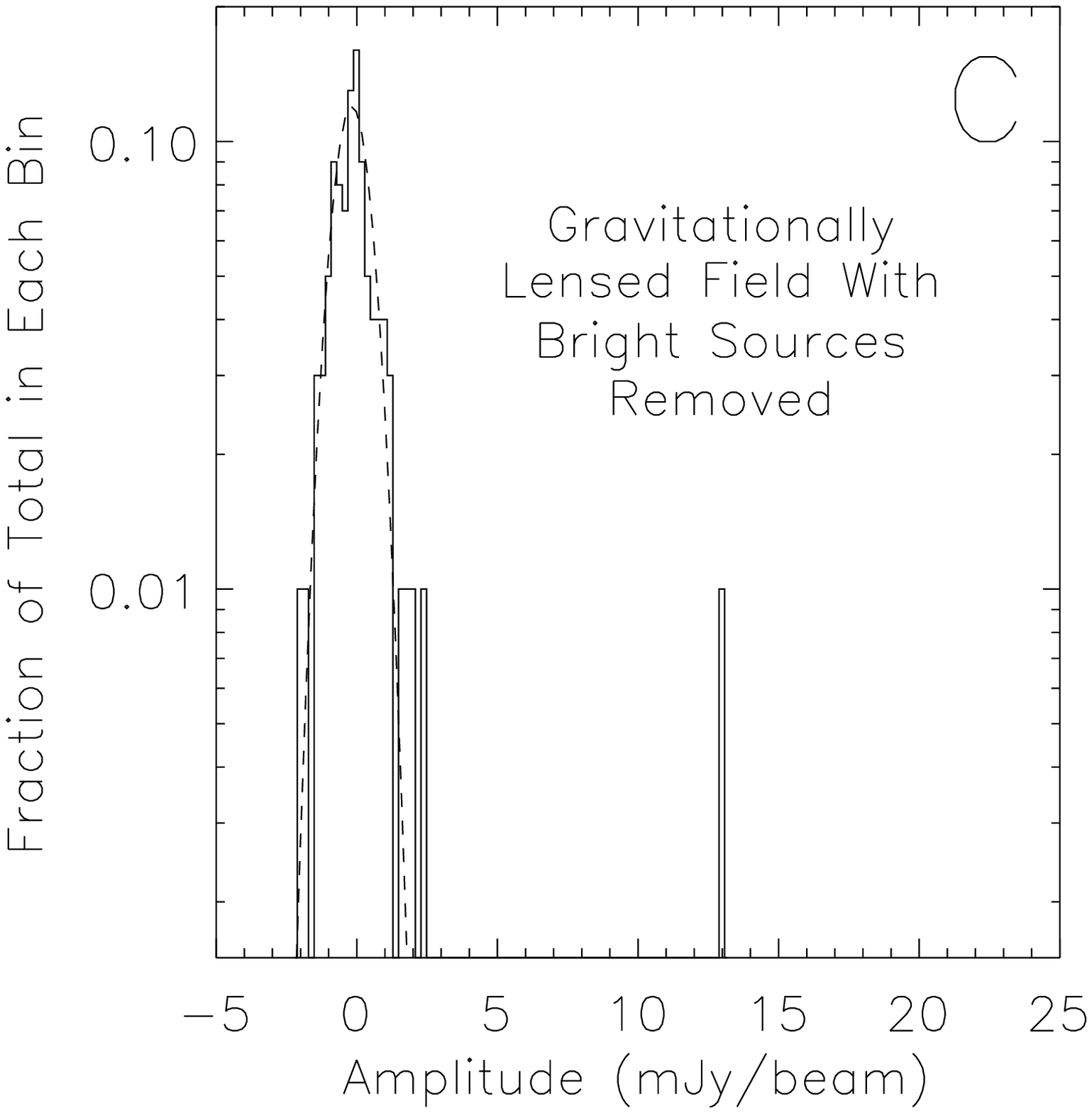,width=0.35\textwidth}
\newline
\caption{Histograms of the amplitudes in the data sets shown in Figure
$2$ (no $\chi^{2}$ cut has been applied).  Gaussian distributions with
mean and variance equal to the final confusion limits quoted in the
text are shown as dashed lines.  Panel A shows isothermal $\beta$
profile fit peak amplitudes for blank sky. Panel B is the same for
gravitationally lensed fields.  Panel C shows amplitudes for
gravitationally lensed fields with sources greater than $6$ mJy
removed.  These show that while source confusion in gravitationally
lensed fields is dominatd by bright sources, confusion in both the
blank sky and bright sources removed cases are dominated by dim
sources.}
\label{fig:3}
\end{figure}

Although not presented as a plot here, it should be noted that
including an SZ increment of amplitude $I_{0}$ mJy per beam in these
simulations does change their character.  The distributions in the
three panels of Figure $2$ are all simply displaced from zero by
$I_{0}$ mJy per beam.

\section{Discussion and Conclusions}

Our simulations show that the SZ effect confusion limit of SCUBA
observations of gravitationally lensed background fields is greater
than that of unlensed fields if no attempt is made to assess quality
of fit or to recognize and remove point sources.

The distributions in Fig.~$3$ can be compared to the results in
Condon (1974) for different values of his parameter $\gamma$, which
parametrizes the brightness of the population of point sources causing
the bulk of the confusion in a data set. The blank field distribution
shown in Fig.~$3$A is well described by a Gaussian distribution
($\gamma \rightarrow 3^{-}$) which shows that unlensed confusion is
dominated by dim sources. Fig.~$3$B shows the distribution from
gravitationally lensed fields; the confusion is dominated by bright
sources ($\gamma \rightarrow 2^{+}$). However, when bright sources are
removed from the gravitationally lensed fields as in Fig.~$3$C, the
confusion limit is reduced to near background levels. This implies
that removing bright point sources allows for a robust measurement of
the SZ increment in galaxy clusters.

Our simulations show that sub-mm sources with fluxes greater than $15$
mJy occur in approximately half of all gravitationally lensed fields.
This matches the count rate of bright sources in sub-mm observations
of galaxy clusters well: about $50$ per cent of clusters observed in
sub-mm surveys contain source brighter than $15$ mJy (e.g. Smail et
al. (2000); Cowie, Barger \& Kneib (2002)).  For comparison, of three
clusters we have observed with SCUBA, one contains an approximately
$40$ mJy gravitationally lensed arc (Scott et al. (2002)), and two do
not contain sources substantially greater than $10$ mJy.

We find that it is possible to detect the SZ increment robustly with
an instrument having sufficient angular resolution that bright sources
can be recognized.  This is because large amplitude gravitationally
lensed background sources produce a signal recognizably different from
the SZ profile.

The confusion limit of cluster fields, once lensed sources have been
removed, is $0.7$ mJy per beam, limiting the signal to noise ratio per
cluster for the thermal effect to be under ten.  A large number of
clusters must be examined in order to obtain a statistically valid
survey of cluster peculiar velocities.  The alternative of observing
only in excellent weather so that SCUBA's short wavelength channel
can aid in identifying confusion sources has not been examined.

We recommend observing a cluster until the 3-$\sigma$ limit for point
source detection is $\sim 6$ mJy.  At this point detector noise
will be a completely inconsequential part of determining the SZ
amplitude; confusion noise will dominate.  Not surprisingly, if point
source confusion and lensing were not an issue one could detect the SZ
increment in much less time.

\section*{Acknowledgments}

Many thanks to Douglas Scott for his helpful suggestions. This work
was supported by the Natural Sciences and Engineering Research Council
of Canada.

\noindent This paper has been produced using the Royal Astronomical
Society/Blackwell Science \LaTeX\ style file.

\end{document}